\documentclass[aps,prl,twocolumn,superscriptaddress]{revtex4-1}
\usepackage{verbatim}
\usepackage{amsmath}
\usepackage[utf8]{inputenc}
\usepackage[pdftex]{graphicx}
\usepackage[separate-uncertainty=true]{siunitx}
\DeclareSIUnit{\rpm}{rpm}
\usepackage[colorlinks=true,urlcolor=blue,linkcolor=blue,citecolor=blue]{hyperref}
\usepackage{xcolor}
\usepackage[T1]{fontenc}
\usepackage{placeins}
\usepackage{nicefrac}
\usepackage{braket}
\usepackage{pdfcomment}
\usepackage{xcolor}
\usepackage{braket}
\usepackage{lineno}
\usepackage{atbegshi}% http://ctan.org/pkg/atbegshi
\newcommand{\HSd}{MoS$_2$/WS$_2$~}

\newcommand{\molyd}{MoS$_2$~}
\newcommand{\moly}{MoS$_2$}
\newcommand{\wolyd}{WS$_2$~}
\newcommand{\woly}{WS$_2$}
\newcommand{\percent}[1]{\SI{#1}{\percent}}

\begin{document}

\date{\today}

\author{Jonas Kiemle}
\affiliation{Walter Schottky Institut and Physics Department, Technical University of Munich, Am Coulombwall 4a, 85748 Garching, Germany}

\author{Florian Sigger}
\affiliation{Walter Schottky Institut and Physics Department, Technical University of Munich, Am Coulombwall 4a, 85748 Garching, Germany}

\author{Michael Lorke}
\affiliation{Bremen Center for Computational Materials Science, University of Bremen, Am Fallturm 1, 28359 Bremen, Germany}
\affiliation{Institute for Theoretical Physics, University of Bremen, 28359 Bremen, Germany}

\author{Bastian Miller}
\affiliation{Walter Schottky Institut and Physics Department, Technical University of Munich, Am Coulombwall 4a, 85748 Garching, Germany}

\author{Kenji Watanabe}
\affiliation{National Institute for Materials Science, Tsukuba, Ibaraki 305-0044, Japan}

\author{Takashi Taniguchi}
\affiliation{National Institute for Materials Science, Tsukuba, Ibaraki 305-0044, Japan}

\author{Alexander Holleitner}
\affiliation{Walter Schottky Institut and Physics Department, Technical University of Munich, Am Coulombwall 4a, 85748 Garching, Germany}
\affiliation{Nanosystems Initiative Munich (NIM), Schellingstr. 4, 80799 München, Germany}

\author{Ursula Wurstbauer}\email{wurstbauer@wwu.de}
\affiliation{Walter Schottky Institut and Physics Department, Technical University of Munich, Am Coulombwall 4a, 85748 Garching, Germany}
\affiliation{Nanosystems Initiative Munich (NIM), Schellingstr. 4, 80799 München, Germany}
\affiliation{Institute of Physics, Westfälische Wilhelms-Universität Münster, Wilhelm-Klemm-Str.10, 48149 Münster, Germany}

\title{Control of the orbital character of indirect excitons in \HSd heterobilayers}

%\begin{document}

%\begin{tocentry}
%\includegraphics{figure_TOC.pdf}
%\end{tocentry}

\begin{abstract}
Valley selective hybridization and residual coupling of electronic states in commensurate van der Waals heterobilayers enable the control of the orbital character of interlayer excitons. We demonstrate electric field control of layer index, orbital character, lifetime and emission energy of indirect excitons in \HSd heterobilayers embedded in an vdW field effect structure. Different excitonic dipoles normal to the layers are found to stem from bound electrons and holes located in different valleys of \HSd with a valley selective degree of hybridization. For the energetically lowest emission lines, coupling of electronic states causes a field-dependent level anticrossing that goes along with a change of the IX lifetime from 400 ns to 100 ns. In the hybridized regime the exiton is delocalized between the two constituent layers, whereas for large positive or negative electric fields, the layer index of the bound hole is field-dependent. Our results demonstrate the design of novel van der Waals solids with the possibility to in-situ control their physical properties via external stimuli such as electric fields.
\end{abstract}


\maketitle
%\newpage

Van der Waals (vdW) heterobilayers (HBs) prepared from optically active transition metal dichalcogenide (TMDC) monolayers, such as \molyd or \woly, combine the excellent properties of the individual layers with the potential for novel functionality provided by the full control of vdW architecture \cite{Geim.2013, Rivera.2018, Alexeev.2019, Jin.2019, Seyler.2019, Tran.2019, Wang.2019}. By stacking two different TMDC monolayers, the electronic bands form an atomically sharp p-n heterojunction, effectively separating photo-generated electron-hole pairs \cite{Zhu.2015b, Chen.2016, Rivera.2018}. This charge transfer allows for the formation of so-called interlayer excitons (IXs) with electrons and holes residing in different layers \cite{Rivera.2015, Rivera.2016}. The reduced overlap of the electron-hole wavefunctions entails greatly prolonged exciton lifetimes of more than \SI{100}{\nano\second} \cite{Miller.2017, Rivera.2015}, exceeding the lifetime of intralayer excitons by several orders of magnitude \cite{Robert.2016, Palummo.2015}. The long lifetimes enable the formation of thermalized dense exciton ensembles \cite{Nagler.2017}, which, together with IX diffusion over several micrometers \cite{Unuchek.2018, Kulig.2018}, sets up the possibility to operate excitonic devices \cite{Unuchek.2018, Jauregui.2018}. 

\begin{figure}
\includegraphics[scale=1]{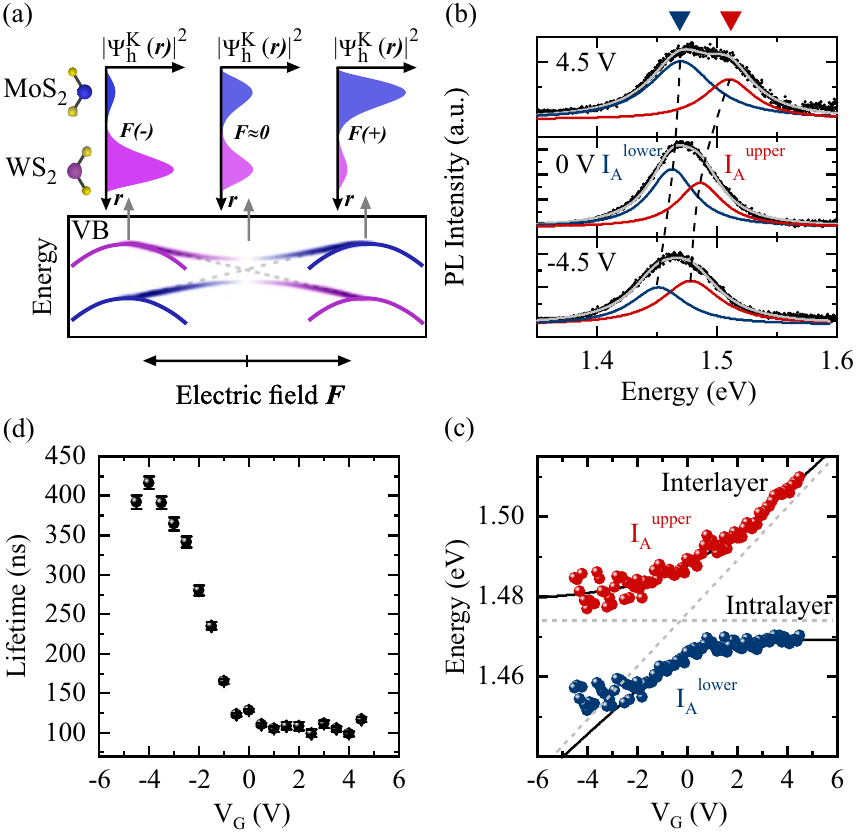}
\caption{(a) Schematic depiction of the squared moduli of the hole wave-functions of the \HSd HB at the topmost VB state at the K point $|\psi_{h}^{K}(\boldsymbol{r})|^{2}$ for large negative $\boldsymbol{F(-)}$, vanishing $\boldsymbol{F} \approx 0$ and large positive electric fields $\boldsymbol{F(+)}$. The coupling of electronic hole states result in a field-tunable hybridization and anti-crossing behavior of the valence band states at the $K^\text{VB}$-point. (b) PL spectra at \SI{10}{\kelvin} for applied voltages $V_{G}$ of \SI{-4.5}{\volt}, \SI{0}{\volt} and \SI{4.5}{\volt}. (c) Emission energy of upper and lower exciton branch as a function of V$_{G}$ indicating the anti-crossing behaviour. The gray dashed lines indicate the interlayer and intralayer character of the exciton emission. (d) PL lifetime of the $I_\text{A}^\text{lower}$ emission line as a function of $V_{G}$, measured at a bath temperature of \SI{10}{\kelvin}. [sample 1]
}
\label{fig: 1}
\end{figure}
Due to hybridization, moiré effects and altered dielectric environment \cite{Gao.2017, Deilmann.2018,Rivera.2018, Alexeev.2019, Jin.2019, Seyler.2019, Tran.2019}, real TMDC HBs can have a much more complicated band structure with fascinating and so far only little explored electronic and optical properties based on the orbital character of the involved electronic states. For the HBs embedded in a field-effect device, an anti-crossing behaviour of IXs is expected based on the residual coupling of electronic states in the two TMDC layers \cite{Gao.2017} accompanied by a gradual change of the exciton nature from primarily interlayer to primarily intralyer [cf. Figure \ref{fig: 1}a]. Since such a coherent coupling occurs across the vdW gap, the field-tunable degree of hybridization and therefore, layer index of electronic states in MoS$_{2}$/WS$_{2}$ HBs opens the door for straight-forward design and fabrication of novel quantum materials, e.g. for the realization of charge qubits in stacked field-effect devices \cite{Lucatto.2019}. Here, we demonstrate the electrical field control of the IXs in MoS$_{2}$/WS$_{2}$ HBs and we find a field-dependent level anti-crossing of the energetically lowest IX emission lines in low-temperature (10K) photoluminescence (PL) [cf. Figure \ref{fig: 1}b,c]. The temperature dependence of the PL in combination with dipole moments from the Stark shift measurements allow the classification of the involved orbital states in MoS$_{2}$/WS$_{2}$ HBs. The observed hybridization effects suggest that vdW HB need to be treated as artificial vdW solids.\par
The relative band alignment of the two layers of the MoS$_{2}$/WS$_{2}$ HBs responds linearly vs an electric field $\boldsymbol{F}$ as $E_\text{\moly}(F)-E_\text{\woly}(F) = E_\text{\moly}(0) - E_\text{\woly}(0) -eFd/\varepsilon$, where $E_\text{\moly}$ and $E_\text{\woly}$ are the band energies of the single layers, $Fd$ describes the voltage drop across the HB and $\varepsilon$ is the effective dielectric function\cite{Gao.2017}. Without finite interlayer coupling, the VB states of the two bands at $K$ would pass through each other. The projection of the electronic wave function on either layer at the reciprocal point $K^\text{VB1}_\text{\woly}$ contains about \percent{10} of \moly, and $K^\text{VB2}_\text{\moly}$ contains \percent{10} of \woly\cite{Gao.2017, Deilmann.2018, SI}. The finite coupling causes level repulsion which manifests in an avoided crossing \cite{Gao.2017} of the IX emission energies [Figure \ref{fig: 1}a]. These VB states bind to electronic states at the conduction band (CB) $\mathit{\Sigma}^\text{CB}_\text{HB}$  that are delocalized between the \molyd (\percent{75}) and \wolyd (\percent{25})\cite{Gao.2017, Deilmann.2018, SI} monolayers forming IXs. The solid line plotted in Figure \ref{fig: 1}d quantifies the coupling strength  $\gamma$ of these IXs using a simplified model of a coupled two-level system
\begin{equation*}
E_\pm =\frac{1}{2}(\Delta(0)-eFd/\varepsilon)\pm\sqrt{(\Delta(0)-eFd/\varepsilon)^2+4\gamma^2},
\end{equation*}
where $E_\pm$ denotes the upper (+) and lower (-) energy branch and $\Delta(0)=E_\text{\moly}(0)-E_\text{\woly}(0)$ \cite{Gao.2017}. The model fit to the data yields $\gamma_+ = \SI{11}{\milli\electronvolt}$ and $\gamma_-=\SI{5}{\milli\electronvolt}$ for the upper and lower branch, respectively. We assume this asymmetry to result from the unequal hybridization at $\mathit{\Sigma}^\text{CB}_\text{HB}$ with \percent{75} \molyd and \percent{25} \woly\cite{Gao.2017, Deilmann.2018, SI}. As a direct consequence of the avoided crossing, the excitons $I_\text{A}^\text{lower}$ gradually change their nature from interlayer, to an intralayer exciton of the artificial van der Waals solid in the anti-crossing regime with delocalized electron at $\mathit{\Sigma}^\text{CB}_\text{HB}$ and delocalized hole at $K^\text{VB}_\text{HB}$. Further increasing the positive gate voltage gradually turns the nature of the exciton back to an IX. This appears analogously for $I^\text{upper}_\text{A}$. In this interpretation, the hole changes its hosting layer, too. For the lowest energy emission line, the hole is localized in \wolyd for large negative gate voltages and localized in \molyd for large positive gate voltages. Due to the altered overlap of the electron and the hole wave-function, we expect also the lifetime of the excitons to depend strongly on the applied voltage $V_{G}$. Indeed, voltage-dependent lifetime measurements of the lowest emission line reveal that the lifetime is about \SI{400}{\nano\second} for negative voltages, where the wave-function overlap should be smallest. The lifetime decreases to about \SI{100}{\nano\second} for zero voltage and saturates for larger voltages as shown in Figure \ref{fig: 1}d. We explain the saturation in lifetime by the similar overlap of electron and hole wave-functions in the hybridized regime (small voltages) and for the hole localized in \molyd (positive voltages).  

\begin{figure}
\includegraphics[scale=1]{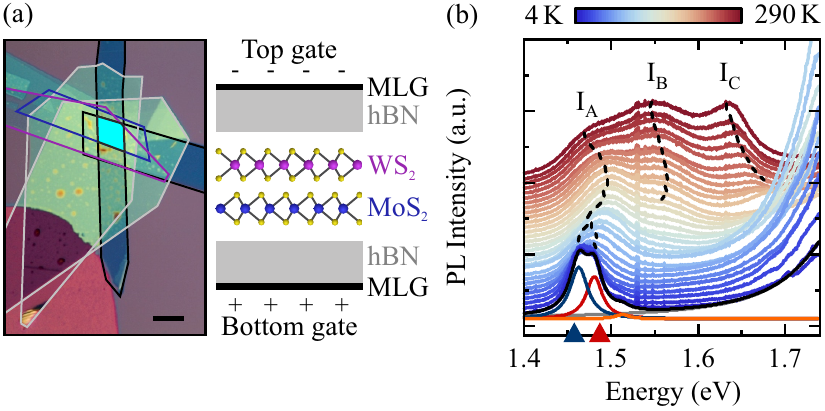}
\caption{(a) Optical microscope image and schematic depiction of the device structure. The turquoise area indicates the HB area. Scale bar is \SI{10}{\micro\meter} [sample 1]. (b) IX PL spectra for bath temperatures between \SI{290}{\kelvin} (dark red) and \SI{4}{\kelvin} (dark blue). The black dashed lines are guides to the eye indicating the emission energy of IXs $I_\text{A}$, $I_\text{B}$ and $I_\text{C}$. [sample 2].
}
\label{fig: 2}
\end{figure}

\begin{figure}
\includegraphics{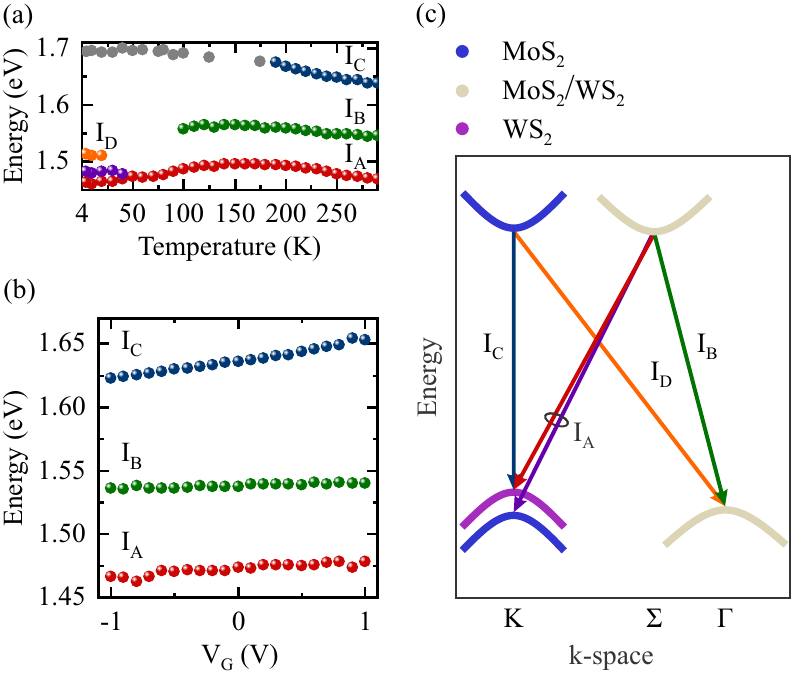}
\caption{(a) Emission energies of IXs $I_\text{A}$, $I_\text{B}$, $I_\text{C}$, and $I_\text{D}$ as a function of bath temperature. [sample 2, grey solid dots of the $I_\text{C}$ branch sample 1]. (b) Voltage-dependent emission energies of IXs at room temperature. [sample 1]. (c) Schematic band structure of a \HSd HB projected onto the atomic orbitals of the individual layers. Blue and purple band colors correspond to \molyd and \wolyd monolayers, while light beige color denotes a large degree of hybridization. The anticipated optical interband transitions are marked by arrows.}
\label{fig: 3}
\end{figure}

The devices in this study consist of \molyd and \wolyd monolayers encapsulated with few-layer hBN and sandwiched between graphite bottom- and top-gate electrodes. The HBs are prepared by micromechanical exfoliation and an all-dry viscoelastic stamping method \cite{CastellanosGomez.2014}. A monolayer \wolyd is stacked onto a monolayer \molyd such that the crystal axes are rotationally aligned to $\ang{0}$ or $\ang{60}$ with a precision of about $\pm\ang{0.5}$ \cite{SI}. A total of 8 hBN-encapsulated \HSd HBs have been fabricated and characterized \cite{SI}. The behaviour of the different devices is consistent and the devices are exceptionally robust against thermal cycling, storage in nitrogen and repeated measurements over months. An optical micrograph and a schematic of the device are shown in Figure \ref{fig: 2}a. Figure \ref{fig: 2}b depicts a temperature series of typical PL spectra taken in a temperature range between room temperature (topmost trace) and 4K (lowest trace). The measurements were carried out using an excitation energy of E$_\text{laser}=\SI{2.54}{\electronvolt}$ and a power of $\text{P}_\text{laser} = \SI{100}{\micro\watt}$, corresponding to about $\SI{10}{\kilo\watt\per\centi\meter^2}$ (spot-size ~$\SI{1}{\micro\meter}$). The emission intensities towards higher energies belong to the direct excitons of \wolyd and \moly [see SI]. The well resolved triplet-structure at higher temperatures occurring between \SI{1.4}{\electronvolt} and \SI{1.7}{\electronvolt} is interpreted to stem from IXs in agreement with a recent report by Okada \textit{et al.} \cite{Okada.2018}. 
For a non-hBN encapsulated \HSd HB, the triplet structure cannot be well resolved \cite{SI} revealing the importance of hBN encapsulation for narrow emission bands \cite{Wierzbowski.2017, Cadiz.2017}. Direct comparison of different samples indicate that the formation of IXs in TMDC HBs does not dependent on the $\ang{0}$ or $\ang{60}$ alignment in agreement with recent reports in literature \cite{Nayak.2017, Okada.2018, Alexeev.2019, SI}. To quantitatively analyze the evolution of the IX emission with temperature the spectra are described by an adequate sum of peak-profiles as exemplary shown for the spectrum at the lowest temperature. The evolution of the IX emission energies with temperature is summarized in Figure \ref{fig: 3}a.
 As expected from an increasing single particle band gap with decreasing temperature, the emission energy of $I_\text{C}$ increases monotonously, similar to the direct excitons (not shown). The emission energy of $I_\text{B}$ is almost constant and the intensity continuously decreases with decreasing temperature, becoming indistinguishable from the noise level below \SI{100}{\kelvin}. The emission energy of $I_\text{A}$ has a very broad maximum around \SI{150}{\kelvin}. The \SI{4}{\kelvin} spectrum in Figure \ref{fig: 2}b clearly demonstrates, that the line splits into a well resolved doublet structure labeled with $I_\text{A}^\text{lower}$ and $I_\text{A}^\text{upper}$ at temperatures below \SI{40}{\kelvin}. Notably, below \SI{30}{\kelvin} another weak emission line $I_\text{D}$ appears energetically just above $I_\text{A}^\text{upper}$. \par

Figure \ref{fig: 3}b depicts the evolution of the peak position of the IX multiplet from Lorentzian fits to the spectra in the presence of an external electric field at room temperature. Therefore a voltage V$_{G}$ is applied between top and bottom graphene electrodes (left axis). Within the considered range of applied voltages, the IX contributions exhibit linear Stark shifts $\Delta E$ of different magnitudes almost independent from the temperature [not shown]. We attribute the non-uniform shifts of the three IX signatures to different out-of-plane dipole moments $\mu_0$. In dependence on the electric field $\boldsymbol{F}$, the Stark shift is given by $\Delta E = \mu_0\boldsymbol{F} + \alpha \boldsymbol{F}^2$, where $\mu_0=ed_0$ is the intrinsic dipole moment, $d_0$ denotes the distance between electron and hole and $\alpha$ is the polarizability of the exciton. Because of the vertical separation of electron and hole, we assume the linear term to be the dominating contribution to the Stark shift. This assumption is in agreement with the experimentally observed linear slopes of the PL peak energies with varying voltage V$_{G}$ [see Figure \ref{fig: 3}b]. The unequal values of $\mu_{0}$ indicate different separations of the electron-hole pairs along the direction of $\boldsymbol{F}$. This is consistent with distinctively different orbital compositions of the electronic states in $k$-space contributing to the excitonic transitions of the \HSd HB. The Stark shifts at room temperature for the three IX transitions are \SI[per-mode=symbol]{15.1}{\milli\electronvolt\per\volt} for $I_\text{C}$, \SI[per-mode=symbol]{2.5}{\milli\electronvolt\per\volt} for $I_\text{B}$, and \SI[per-mode=symbol]{6.3}{\milli\electronvolt\per\volt} for $I_\text{A}$. The estimated separation of the electron and hole along the electric field constitute approximately \SI{1.8}{\angstrom},  \SI{0.3}{\angstrom} and \SI{0.8}{\angstrom} for $I_\text{C}$, $I_\text{B}$ and $I_\text{A}$, respectively. The low temperature values for the stark shifts $\Delta E$ at \SI{4}{\kelvin} are very similar to the values at room temperatures and are \SI[per-mode=symbol]{17.0}{\milli\electronvolt\per\volt} for $I_\text{C}$ and $\approx\SI[per-mode=symbol]{7} {\milli\electronvolt\per\volt}$ for $I_\text{A}^\text{lower}$, $I_\text{A}^\text{upper}$ and $I_\text{D}$, respectively. 
\par

The peculiar temperature dependence together with the significantly different Stark shifts strongly suggest that the IXs are formed by electrons and holes residing in different valleys of the BZ. The conduction (CB) and valence bands (VB) of the \HSd HB near the relevant critical points at the $\mathit{\Gamma}$, $K$ and $\mathit{\Sigma}$ valleys are sketched in Figure \ref{fig: 3}c following recent theoretical work \cite{Gao.2017, Deilmann.2018, SI}. The single particle bands are projected onto the atomic orbitals of the individual layers that are reflected by the color code. Red and blue colors correspond to electronic states purely from \molyd and \woly, respectively, whereas light beige indicates hybridization of states from both layers. The hybridization is a consequence of coupling between electronic states on the molecular-orbital level that is particularly strong at the $\mathit{\Sigma}$ valley in the CB and the $\mathit{\Gamma}$ valley in the VB. From DFT calculations, the relevant high symmetry points of the band structure appear to be similar for AB and AA stacked \HSd HBs \cite{SI}.\par

The lowest CB states at the $K^\text{CB}$-point are of \percent{100} \molyd character, whereas the two topmost VB states at the $K^\text{VB}$-point are split by only $\approx \SI{16}{\milli\electronvolt}$ and are of inverse orbital character with around \percent{90} \wolyd and \percent{90} \molyd for the higher and lower state \cite{SI}. Similar to TMDC homo-bilayers, also the \HSd HBs are indirect semiconductors with the lowest energy transition between the  $\mathit{\Sigma}^\text{CB}$ and the $K^\text{VB}$ valley \cite{Deilmann.2018}.\par
Due to of the strong hybridization of the electronic states for some high symmetry points, those electrons and holes are delocalized between the two layers and, hence, behave more like charge carriers of a new `artificial vdW solid' than of a vdW HB. Guided by this aspect, we assign the emission line of $I_\text{B}$ with smallest Stark shift and consequently the smallest separation between electron-hole pairs, to be formed by an electron at $\mathit{\Sigma}^\text{CB}_\text{HB}$ and a hole at $\mathit{\Gamma}^\text{VB}_\text{HB}$. The small but finite vertical separation of the charge carriers results from a large degree of orbital hybridization with approx. \percent{75} \molyd character at the $\mathit{\Sigma}^\text{CB}$-point and approx. \percent{60} \wolyd character at the $\mathit{\Gamma}^\text{VB}$ point\cite{Gao.2017, SI}. Therefore, $I_\text{B}$ can be interpreted as momentum indirect, but real space more direct transition of the \HSd vdW solid.
With similar arguments, we assign $I_\text{C}$, which shows the largest Stark shift to be an IX with the electron in the fully layer polarized $K^\text{CB}_\text{\moly}$-point ($\approx$ \percent{100} \moly \cite{Gao.2017}) and the hole in the almost fully layer polarized $K^\text{VB1}_\text{\woly}$-point ($\approx$ \percent{90} \woly \cite{Gao.2017}). The remaining excitons $I_\text{A}$ and $I_\text{D}$ exhibit all similar Stark shift values that are between those of $I_\text{C}$ and $I_\text{B}$. For this reason, it is most likely that for these interband transitions one of the charge carriers is more delocalized between the two layers and resides in a highly hybridized valley, while the other coulomb coupled charge carrier is localized in either \molyd or \wolyd and belongs therefore to a non- or only weakly hybridized valley.\par
Taking into account the energies of the IXs with  $E(I_\text{D}) > E(I_\text{A}^\text{upper})> E(I_\text{A}^\text{lower})$, we assign $I_\text{D}$ to the transition with the electron at $K^\text{CB}_\text{\moly}$ and the hole at the hybridized $\mathit{\Gamma}^\text{VB}_\text{HB}$-point. The lowest energy emission lines $I_\text{A}^\text{lower}$ and $I_\text{A}^\text{upper}$ are those showing anti-crossing behaviour in electric field-dependent measurements.
The experimentally found energy separation of the two emission lines is about \SI{20}{\milli\electronvolt} which is in excellent agreement with the theoretical value of about \SI{16}{\milli\electronvolt} obtained from DFT calculation \cite{SI}. In brief, the five red-shifted emission lines are assigned to a momentum direct interlayer transition $I_\text{C}$ from $K^\text{CB}_\text{\moly}$ to $K^\text{VB1}_\text{\woly}$, a  momentum indirect intralayer transition $I_\text{B}$ from $\mathit{\Sigma}^\text{CV}_\text{HB}$ to $\mathit{\Gamma}^\text{VB}_\text{HB}$ and three momentum indirect interlayer transitions, namely $I_\text{A}^\text{lower}$ from $\mathit{\Sigma}^\text{CB}_\text{HB}$ to  $K^\text{VB1}_\text{\woly}$, $I_\text{A}^\text{upper}$ from $\mathit{\Sigma}^\text{CB}_\text{HB}$ to $K^\text{VB2}_\text{\moly}$ and $I_\text{D}$ from $K^\text{CB}_\text{\moly}$ to $\mathit{\Gamma}^\text{VB}_\text{HB}$.\par
%The calculated band structure suggests that the interband transition energy referring to $I_\text{B}$ is smaller than \cite{Gao.2017} or almost equal \cite{Deilmann.2018} to the transition energies of $I_\text{A}^\text{lower}$ and $I_\text{A}^\text{upper}$ in disagreement to our assignment based on Stark shift arguments. This discrepancy can be obliterated considering different exciton binding energies $E_\text{exc}$ lowering the optically measured transition energies. The exciton binding energy $E_\text{exc}$ is inversely proportional to the reduced mass $\mu$ of the exciton with $1/\mu = 1/m_\text{e} + 1/m_\text{h}$, where $m_\text{e(h)}$ denotes the effective electron (hole) mass. By estimating the relative effective masses from the reported band structure \cite{Gao.2017,Deilmann.2018}, we find that $E_\text{exc}(I_\text{C}) > E_\text{exc}(I_\text{D}) > E_\text{exc}(I_\text{A}) > E_\text{exc}(I_\text{B})$ indicating that the emission energy of $I_\text{B}$ can indeed be larger than that of $I_\text{A}^\text{lower}$ and $I_\text{A}^\text{upper}$ due to the lower exciton binding energy.\par
The increase of the emission energy of $I_\text{C}$ with decreasing temperature between \SI{290}{\kelvin} and \SI{150}{\kelvin} is consistent with an increase of the single particle band gap of both constituent layers \cite{Varshni.1967}. The subsequent lowering of the emission energy below \SI{150}{\kelvin} and reduced emission intensity might be caused by a reduction of the amount of thermally excited kinematic excitons with $|\Delta k| > 0$ that can couple to the light, since we expect a slight lattice mismatch to cause a slight misalignment of the \molyd  and \wolyd  $K$-points \cite{Miller.2017, Yu.2015, SI}. The slight mismatch in the momenta at this quasi direct interband transition at the $K$-points  might also be the reason for the reduced emission intensity with reduced temperature. The phonon-activated transition $I_\text{B}$ is rather broad and disappears below about \SI{100}{\kelvin}, which is expected to be caused by the combined effect of reduced electron-phonon coupling and freezing of the relevant phonon modes \cite{Ge.2013}.\par
To conclude, we demonstrate widely tunable IX emission by applying an electric field perpendicular to the HB plane. The results are interpreted in terms of finite interlayer coupling, resulting in different orbital hybridization dominating the multiphysics of IX transitions in \HSd HBs. Most remarkably, the field-tunable anti-crossing behaviour of the lowest IX emission lines goes along with a field-dependent change of the exciton character from interlayer to intralayer and back to interlayer, accompanied with a transfer of the participating hole between the constituent layers.  These types of HBs enable the realization of a new type of charge qubits \cite{Lucatto.2019}. The observed valley-dependent hybridization effects manifest that vdW HBs need to be treated as artificial vdW solids.\par
We gratefully acknowledge financial support by the Deutsche Forschungsgemeinschaft (DFG) via excellence cluster `Nanosystems Initiative Munich' (NIM) and DFG projects WU 637/4- 1 and HO 3324/9-1. M.L. was supported by the Deutsche Forschungsgemeinschaft (DFG) within RTG 2247 and through a grant for CPU time at the HLRN (Hannover/Berlin). K.W. and T.T. acknowledge support from the Elemental Strategy Initiative conducted by the MEXT, Japan and and the CREST (JPMJCR15F3), JST.

%\subsection{Supplementary Information} accompanies this paper

%\subsection{Competing financial interests} The authors declare no competing financial interests.

%\clearpage
%\newpage

\bibliography{main_revised}

\begin{thebibliography}{30}%
\makeatletter
\providecommand \@ifxundefined [1]{%
 \@ifx{#1\undefined}
}%
\providecommand \@ifnum [1]{%
 \ifnum #1\expandafter \@firstoftwo
 \else \expandafter \@secondoftwo
 \fi
}%
\providecommand \@ifx [1]{%
 \ifx #1\expandafter \@firstoftwo
 \else \expandafter \@secondoftwo
 \fi
}%
\providecommand \natexlab [1]{#1}%
\providecommand \enquote  [1]{``#1''}%
\providecommand \bibnamefont  [1]{#1}%
\providecommand \bibfnamefont [1]{#1}%
\providecommand \citenamefont [1]{#1}%
\providecommand \href@noop [0]{\@secondoftwo}%
\providecommand \href [0]{\begingroup \@sanitize@url \@href}%
\providecommand \@href[1]{\@@startlink{#1}\@@href}%
\providecommand \@@href[1]{\endgroup#1\@@endlink}%
\providecommand \@sanitize@url [0]{\catcode `\\12\catcode `\$12\catcode
  `\&12\catcode `\#12\catcode `\^12\catcode `\_12\catcode `\%12\relax}%
\providecommand \@@startlink[1]{}%
\providecommand \@@endlink[0]{}%
\providecommand \url  [0]{\begingroup\@sanitize@url \@url }%
\providecommand \@url [1]{\endgroup\@href {#1}{\urlprefix }}%
\providecommand \urlprefix  [0]{URL }%
\providecommand \Eprint [0]{\href }%
\providecommand \doibase [0]{http://dx.doi.org/}%
\providecommand \selectlanguage [0]{\@gobble}%
\providecommand \bibinfo  [0]{\@secondoftwo}%
\providecommand \bibfield  [0]{\@secondoftwo}%
\providecommand \translation [1]{[#1]}%
\providecommand \BibitemOpen [0]{}%
\providecommand \bibitemStop [0]{}%
\providecommand \bibitemNoStop [0]{.\EOS\space}%
\providecommand \EOS [0]{\spacefactor3000\relax}%
\providecommand \BibitemShut  [1]{\csname bibitem#1\endcsname}%
\let\auto@bib@innerbib\@empty
%</preamble>
\bibitem [{\citenamefont {Geim}\ and\ \citenamefont
  {Grigorieva}(2013)}]{Geim.2013}%
  \BibitemOpen
  \bibfield  {author} {\bibinfo {author} {\bibfnamefont {A.~K.}\ \bibnamefont
  {Geim}}\ and\ \bibinfo {author} {\bibfnamefont {I.~V.}\ \bibnamefont
  {Grigorieva}},\ }\href {\doibase 10.1038/nature12385} {\bibfield  {journal}
  {\bibinfo  {journal} {Nature}\ }\textbf {\bibinfo {volume} {499}},\ \bibinfo
  {pages} {419} (\bibinfo {year} {2013})}\BibitemShut {NoStop}%
\bibitem [{\citenamefont {Rivera}\ \emph {et~al.}(2018)\citenamefont {Rivera},
  \citenamefont {Yu}, \citenamefont {Seyler}, \citenamefont {Wilson},
  \citenamefont {Yao},\ and\ \citenamefont {Xu}}]{Rivera.2018}%
  \BibitemOpen
  \bibfield  {author} {\bibinfo {author} {\bibfnamefont {P.}~\bibnamefont
  {Rivera}}, \bibinfo {author} {\bibfnamefont {H.}~\bibnamefont {Yu}}, \bibinfo
  {author} {\bibfnamefont {K.~L.}\ \bibnamefont {Seyler}}, \bibinfo {author}
  {\bibfnamefont {N.~P.}\ \bibnamefont {Wilson}}, \bibinfo {author}
  {\bibfnamefont {W.}~\bibnamefont {Yao}}, \ and\ \bibinfo {author}
  {\bibfnamefont {X.}~\bibnamefont {Xu}},\ }\href {\doibase
  10.1038/s41565-018-0193-0} {\bibfield  {journal} {\bibinfo  {journal} {Nature
  nanotechnology}\ }\textbf {\bibinfo {volume} {13}},\ \bibinfo {pages} {1004}
  (\bibinfo {year} {2018})}\BibitemShut {NoStop}%
\bibitem [{\citenamefont {Alexeev}\ \emph {et~al.}(2019)\citenamefont
  {Alexeev}, \citenamefont {Ruiz-Tijerina}, \citenamefont {Danovich},
  \citenamefont {Hamer}, \citenamefont {Terry}, \citenamefont {Nayak},
  \citenamefont {Ahn}, \citenamefont {Pak}, \citenamefont {Lee}, \citenamefont
  {Sohn}, \citenamefont {Molas}, \citenamefont {Koperski}, \citenamefont
  {Watanabe}, \citenamefont {Taniguchi}, \citenamefont {Novoselov},
  \citenamefont {Gorbachev}, \citenamefont {Shin}, \citenamefont {Fal'ko},\
  and\ \citenamefont {Tartakovskii}}]{Alexeev.2019}%
  \BibitemOpen
  \bibfield  {author} {\bibinfo {author} {\bibfnamefont {E.~M.}\ \bibnamefont
  {Alexeev}}, \bibinfo {author} {\bibfnamefont {D.~A.}\ \bibnamefont
  {Ruiz-Tijerina}}, \bibinfo {author} {\bibfnamefont {M.}~\bibnamefont
  {Danovich}}, \bibinfo {author} {\bibfnamefont {M.~J.}\ \bibnamefont {Hamer}},
  \bibinfo {author} {\bibfnamefont {D.~J.}\ \bibnamefont {Terry}}, \bibinfo
  {author} {\bibfnamefont {P.~K.}\ \bibnamefont {Nayak}}, \bibinfo {author}
  {\bibfnamefont {S.}~\bibnamefont {Ahn}}, \bibinfo {author} {\bibfnamefont
  {S.}~\bibnamefont {Pak}}, \bibinfo {author} {\bibfnamefont {J.}~\bibnamefont
  {Lee}}, \bibinfo {author} {\bibfnamefont {J.~I.}\ \bibnamefont {Sohn}},
  \bibinfo {author} {\bibfnamefont {M.~R.}\ \bibnamefont {Molas}}, \bibinfo
  {author} {\bibfnamefont {M.}~\bibnamefont {Koperski}}, \bibinfo {author}
  {\bibfnamefont {K.}~\bibnamefont {Watanabe}}, \bibinfo {author}
  {\bibfnamefont {T.}~\bibnamefont {Taniguchi}}, \bibinfo {author}
  {\bibfnamefont {K.~S.}\ \bibnamefont {Novoselov}}, \bibinfo {author}
  {\bibfnamefont {R.~V.}\ \bibnamefont {Gorbachev}}, \bibinfo {author}
  {\bibfnamefont {H.~S.}\ \bibnamefont {Shin}}, \bibinfo {author}
  {\bibfnamefont {V.~I.}\ \bibnamefont {Fal'ko}}, \ and\ \bibinfo {author}
  {\bibfnamefont {A.~I.}\ \bibnamefont {Tartakovskii}},\ }\href {\doibase
  10.1038/s41586-019-0986-9} {\bibfield  {journal} {\bibinfo  {journal}
  {Nature}\ }\textbf {\bibinfo {volume} {567}},\ \bibinfo {pages} {81}
  (\bibinfo {year} {2019})}\BibitemShut {NoStop}%
\bibitem [{\citenamefont {Jin}\ \emph {et~al.}(2019)\citenamefont {Jin},
  \citenamefont {Regan}, \citenamefont {Yan}, \citenamefont {{Iqbal Bakti
  Utama}}, \citenamefont {Wang}, \citenamefont {Zhao}, \citenamefont {Qin},
  \citenamefont {Yang}, \citenamefont {Zheng}, \citenamefont {Shi},
  \citenamefont {Watanabe}, \citenamefont {Taniguchi}, \citenamefont {Tongay},
  \citenamefont {Zettl},\ and\ \citenamefont {Wang}}]{Jin.2019}%
  \BibitemOpen
  \bibfield  {author} {\bibinfo {author} {\bibfnamefont {C.}~\bibnamefont
  {Jin}}, \bibinfo {author} {\bibfnamefont {E.~C.}\ \bibnamefont {Regan}},
  \bibinfo {author} {\bibfnamefont {A.}~\bibnamefont {Yan}}, \bibinfo {author}
  {\bibfnamefont {M.}~\bibnamefont {{Iqbal Bakti Utama}}}, \bibinfo {author}
  {\bibfnamefont {D.}~\bibnamefont {Wang}}, \bibinfo {author} {\bibfnamefont
  {S.}~\bibnamefont {Zhao}}, \bibinfo {author} {\bibfnamefont {Y.}~\bibnamefont
  {Qin}}, \bibinfo {author} {\bibfnamefont {S.}~\bibnamefont {Yang}}, \bibinfo
  {author} {\bibfnamefont {Z.}~\bibnamefont {Zheng}}, \bibinfo {author}
  {\bibfnamefont {S.}~\bibnamefont {Shi}}, \bibinfo {author} {\bibfnamefont
  {K.}~\bibnamefont {Watanabe}}, \bibinfo {author} {\bibfnamefont
  {T.}~\bibnamefont {Taniguchi}}, \bibinfo {author} {\bibfnamefont
  {S.}~\bibnamefont {Tongay}}, \bibinfo {author} {\bibfnamefont
  {A.}~\bibnamefont {Zettl}}, \ and\ \bibinfo {author} {\bibfnamefont
  {F.}~\bibnamefont {Wang}},\ }\href {\doibase 10.1038/s41586-019-0976-y}
  {\bibfield  {journal} {\bibinfo  {journal} {Nature}\ }\textbf {\bibinfo
  {volume} {567}},\ \bibinfo {pages} {76} (\bibinfo {year} {2019})}\BibitemShut
  {NoStop}%
\bibitem [{\citenamefont {Seyler}\ \emph {et~al.}(2019)\citenamefont {Seyler},
  \citenamefont {Rivera}, \citenamefont {Yu}, \citenamefont {Wilson},
  \citenamefont {Ray}, \citenamefont {Mandrus}, \citenamefont {Yan},
  \citenamefont {Yao},\ and\ \citenamefont {Xu}}]{Seyler.2019}%
  \BibitemOpen
  \bibfield  {author} {\bibinfo {author} {\bibfnamefont {K.~L.}\ \bibnamefont
  {Seyler}}, \bibinfo {author} {\bibfnamefont {P.}~\bibnamefont {Rivera}},
  \bibinfo {author} {\bibfnamefont {H.}~\bibnamefont {Yu}}, \bibinfo {author}
  {\bibfnamefont {N.~P.}\ \bibnamefont {Wilson}}, \bibinfo {author}
  {\bibfnamefont {E.~L.}\ \bibnamefont {Ray}}, \bibinfo {author} {\bibfnamefont
  {D.~G.}\ \bibnamefont {Mandrus}}, \bibinfo {author} {\bibfnamefont
  {J.}~\bibnamefont {Yan}}, \bibinfo {author} {\bibfnamefont {W.}~\bibnamefont
  {Yao}}, \ and\ \bibinfo {author} {\bibfnamefont {X.}~\bibnamefont {Xu}},\
  }\href {\doibase 10.1038/s41586-019-0957-1} {\bibfield  {journal} {\bibinfo
  {journal} {Nature}\ }\textbf {\bibinfo {volume} {567}},\ \bibinfo {pages}
  {66} (\bibinfo {year} {2019})}\BibitemShut {NoStop}%
\bibitem [{\citenamefont {Tran}\ \emph {et~al.}(2019)\citenamefont {Tran},
  \citenamefont {Moody}, \citenamefont {Wu}, \citenamefont {Lu}, \citenamefont
  {Choi}, \citenamefont {Kim}, \citenamefont {Rai}, \citenamefont {Sanchez},
  \citenamefont {Quan}, \citenamefont {Singh}, \citenamefont {Embley},
  \citenamefont {Zepeda}, \citenamefont {Campbell}, \citenamefont {Autry},
  \citenamefont {Taniguchi}, \citenamefont {Watanabe}, \citenamefont {Lu},
  \citenamefont {Banerjee}, \citenamefont {Silverman}, \citenamefont {Kim},
  \citenamefont {Tutuc}, \citenamefont {Yang}, \citenamefont {Macdonald},\ and\
  \citenamefont {Li}}]{Tran.2019}%
  \BibitemOpen
  \bibfield  {author} {\bibinfo {author} {\bibfnamefont {K.}~\bibnamefont
  {Tran}}, \bibinfo {author} {\bibfnamefont {G.}~\bibnamefont {Moody}},
  \bibinfo {author} {\bibfnamefont {F.}~\bibnamefont {Wu}}, \bibinfo {author}
  {\bibfnamefont {X.}~\bibnamefont {Lu}}, \bibinfo {author} {\bibfnamefont
  {J.}~\bibnamefont {Choi}}, \bibinfo {author} {\bibfnamefont {K.}~\bibnamefont
  {Kim}}, \bibinfo {author} {\bibfnamefont {A.}~\bibnamefont {Rai}}, \bibinfo
  {author} {\bibfnamefont {D.~A.}\ \bibnamefont {Sanchez}}, \bibinfo {author}
  {\bibfnamefont {J.}~\bibnamefont {Quan}}, \bibinfo {author} {\bibfnamefont
  {A.}~\bibnamefont {Singh}}, \bibinfo {author} {\bibfnamefont
  {J.}~\bibnamefont {Embley}}, \bibinfo {author} {\bibfnamefont
  {A.}~\bibnamefont {Zepeda}}, \bibinfo {author} {\bibfnamefont
  {M.}~\bibnamefont {Campbell}}, \bibinfo {author} {\bibfnamefont
  {T.}~\bibnamefont {Autry}}, \bibinfo {author} {\bibfnamefont
  {T.}~\bibnamefont {Taniguchi}}, \bibinfo {author} {\bibfnamefont
  {K.}~\bibnamefont {Watanabe}}, \bibinfo {author} {\bibfnamefont
  {N.}~\bibnamefont {Lu}}, \bibinfo {author} {\bibfnamefont {S.~K.}\
  \bibnamefont {Banerjee}}, \bibinfo {author} {\bibfnamefont {K.~L.}\
  \bibnamefont {Silverman}}, \bibinfo {author} {\bibfnamefont {S.}~\bibnamefont
  {Kim}}, \bibinfo {author} {\bibfnamefont {E.}~\bibnamefont {Tutuc}}, \bibinfo
  {author} {\bibfnamefont {L.}~\bibnamefont {Yang}}, \bibinfo {author}
  {\bibfnamefont {A.~H.}\ \bibnamefont {Macdonald}}, \ and\ \bibinfo {author}
  {\bibfnamefont {X.}~\bibnamefont {Li}},\ }\href {\doibase
  10.1038/s41586-019-0975-z} {\bibfield  {journal} {\bibinfo  {journal}
  {Nature}\ }\textbf {\bibinfo {volume} {567}},\ \bibinfo {pages} {71}
  (\bibinfo {year} {2019})}\BibitemShut {NoStop}%
\bibitem [{\citenamefont {Wang}\ \emph {et~al.}(2019)\citenamefont {Wang},
  \citenamefont {Rhodes}, \citenamefont {Watanabe}, \citenamefont {Taniguchi},
  \citenamefont {Hone}, \citenamefont {Shan},\ and\ \citenamefont
  {Mak}}]{Wang.2019}%
  \BibitemOpen
  \bibfield  {author} {\bibinfo {author} {\bibfnamefont {Z.}~\bibnamefont
  {Wang}}, \bibinfo {author} {\bibfnamefont {D.~A.}\ \bibnamefont {Rhodes}},
  \bibinfo {author} {\bibfnamefont {K.}~\bibnamefont {Watanabe}}, \bibinfo
  {author} {\bibfnamefont {T.}~\bibnamefont {Taniguchi}}, \bibinfo {author}
  {\bibfnamefont {J.~C.}\ \bibnamefont {Hone}}, \bibinfo {author}
  {\bibfnamefont {J.}~\bibnamefont {Shan}}, \ and\ \bibinfo {author}
  {\bibfnamefont {K.~F.}\ \bibnamefont {Mak}},\ }\href {\doibase
  10.1038/s41586-019-1591-7} {\bibfield  {journal} {\bibinfo  {journal}
  {Nature}\ }\textbf {\bibinfo {volume} {574}},\ \bibinfo {pages} {76}
  (\bibinfo {year} {2019})}\BibitemShut {NoStop}%
\bibitem [{\citenamefont {Zhu}\ \emph {et~al.}(2015)\citenamefont {Zhu},
  \citenamefont {Monahan}, \citenamefont {Gong}, \citenamefont {Zhu},
  \citenamefont {Williams},\ and\ \citenamefont {Nelson}}]{Zhu.2015b}%
  \BibitemOpen
  \bibfield  {author} {\bibinfo {author} {\bibfnamefont {X.}~\bibnamefont
  {Zhu}}, \bibinfo {author} {\bibfnamefont {N.~R.}\ \bibnamefont {Monahan}},
  \bibinfo {author} {\bibfnamefont {Z.}~\bibnamefont {Gong}}, \bibinfo {author}
  {\bibfnamefont {H.}~\bibnamefont {Zhu}}, \bibinfo {author} {\bibfnamefont
  {K.~W.}\ \bibnamefont {Williams}}, \ and\ \bibinfo {author} {\bibfnamefont
  {C.~A.}\ \bibnamefont {Nelson}},\ }\href {\doibase 10.1021/jacs.5b03141}
  {\bibfield  {journal} {\bibinfo  {journal} {Journal of the American Chemical
  Society}\ }\textbf {\bibinfo {volume} {137}},\ \bibinfo {pages} {8313}
  (\bibinfo {year} {2015})}\BibitemShut {NoStop}%
\bibitem [{\citenamefont {Chen}\ \emph {et~al.}(2016)\citenamefont {Chen},
  \citenamefont {Wen}, \citenamefont {Zhang}, \citenamefont {Wu}, \citenamefont
  {Gong}, \citenamefont {Zhang}, \citenamefont {Yuan}, \citenamefont {Yi},
  \citenamefont {Lou}, \citenamefont {Ajayan}, \citenamefont {Zhuang},
  \citenamefont {Zhang},\ and\ \citenamefont {Zheng}}]{Chen.2016}%
  \BibitemOpen
  \bibfield  {author} {\bibinfo {author} {\bibfnamefont {H.}~\bibnamefont
  {Chen}}, \bibinfo {author} {\bibfnamefont {X.}~\bibnamefont {Wen}}, \bibinfo
  {author} {\bibfnamefont {J.}~\bibnamefont {Zhang}}, \bibinfo {author}
  {\bibfnamefont {T.}~\bibnamefont {Wu}}, \bibinfo {author} {\bibfnamefont
  {Y.}~\bibnamefont {Gong}}, \bibinfo {author} {\bibfnamefont {X.}~\bibnamefont
  {Zhang}}, \bibinfo {author} {\bibfnamefont {J.}~\bibnamefont {Yuan}},
  \bibinfo {author} {\bibfnamefont {C.}~\bibnamefont {Yi}}, \bibinfo {author}
  {\bibfnamefont {J.}~\bibnamefont {Lou}}, \bibinfo {author} {\bibfnamefont
  {P.~M.}\ \bibnamefont {Ajayan}}, \bibinfo {author} {\bibfnamefont
  {W.}~\bibnamefont {Zhuang}}, \bibinfo {author} {\bibfnamefont
  {G.}~\bibnamefont {Zhang}}, \ and\ \bibinfo {author} {\bibfnamefont
  {J.}~\bibnamefont {Zheng}},\ }\href {\doibase 10.1038/ncomms12512} {\bibfield
   {journal} {\bibinfo  {journal} {Nature communications}\ }\textbf {\bibinfo
  {volume} {7}},\ \bibinfo {pages} {12512} (\bibinfo {year}
  {2016})}\BibitemShut {NoStop}%
\bibitem [{\citenamefont {Rivera}\ \emph {et~al.}(2015)\citenamefont {Rivera},
  \citenamefont {Schaibley}, \citenamefont {Jones}, \citenamefont {Ross},
  \citenamefont {Wu}, \citenamefont {Aivazian}, \citenamefont {Klement},
  \citenamefont {Seyler}, \citenamefont {Clark}, \citenamefont {Ghimire},
  \citenamefont {Yan}, \citenamefont {Mandrus}, \citenamefont {Yao},\ and\
  \citenamefont {Xu}}]{Rivera.2015}%
  \BibitemOpen
  \bibfield  {author} {\bibinfo {author} {\bibfnamefont {P.}~\bibnamefont
  {Rivera}}, \bibinfo {author} {\bibfnamefont {J.~R.}\ \bibnamefont
  {Schaibley}}, \bibinfo {author} {\bibfnamefont {A.~M.}\ \bibnamefont
  {Jones}}, \bibinfo {author} {\bibfnamefont {J.~S.}\ \bibnamefont {Ross}},
  \bibinfo {author} {\bibfnamefont {S.}~\bibnamefont {Wu}}, \bibinfo {author}
  {\bibfnamefont {G.}~\bibnamefont {Aivazian}}, \bibinfo {author}
  {\bibfnamefont {P.}~\bibnamefont {Klement}}, \bibinfo {author} {\bibfnamefont
  {K.}~\bibnamefont {Seyler}}, \bibinfo {author} {\bibfnamefont
  {G.}~\bibnamefont {Clark}}, \bibinfo {author} {\bibfnamefont {N.~J.}\
  \bibnamefont {Ghimire}}, \bibinfo {author} {\bibfnamefont {J.}~\bibnamefont
  {Yan}}, \bibinfo {author} {\bibfnamefont {D.~G.}\ \bibnamefont {Mandrus}},
  \bibinfo {author} {\bibfnamefont {W.}~\bibnamefont {Yao}}, \ and\ \bibinfo
  {author} {\bibfnamefont {X.}~\bibnamefont {Xu}},\ }\href {\doibase
  10.1038/ncomms7242} {\bibfield  {journal} {\bibinfo  {journal} {Nature
  communications}\ }\textbf {\bibinfo {volume} {6}},\ \bibinfo {pages} {6242}
  (\bibinfo {year} {2015})}\BibitemShut {NoStop}%
\bibitem [{\citenamefont {Rivera}\ \emph {et~al.}(2016)\citenamefont {Rivera},
  \citenamefont {Seyler}, \citenamefont {Yu}, \citenamefont {Schaibley},
  \citenamefont {Yan}, \citenamefont {Mandrus}, \citenamefont {Yao},\ and\
  \citenamefont {Xu}}]{Rivera.2016}%
  \BibitemOpen
  \bibfield  {author} {\bibinfo {author} {\bibfnamefont {P.}~\bibnamefont
  {Rivera}}, \bibinfo {author} {\bibfnamefont {K.~L.}\ \bibnamefont {Seyler}},
  \bibinfo {author} {\bibfnamefont {H.}~\bibnamefont {Yu}}, \bibinfo {author}
  {\bibfnamefont {J.~R.}\ \bibnamefont {Schaibley}}, \bibinfo {author}
  {\bibfnamefont {J.}~\bibnamefont {Yan}}, \bibinfo {author} {\bibfnamefont
  {D.~G.}\ \bibnamefont {Mandrus}}, \bibinfo {author} {\bibfnamefont
  {W.}~\bibnamefont {Yao}}, \ and\ \bibinfo {author} {\bibfnamefont
  {X.}~\bibnamefont {Xu}},\ }\href {\doibase 10.1126/science.aac7820}
  {\bibfield  {journal} {\bibinfo  {journal} {Science}\ }\textbf {\bibinfo
  {volume} {351}},\ \bibinfo {pages} {688} (\bibinfo {year}
  {2016})}\BibitemShut {NoStop}%
\bibitem [{\citenamefont {Miller}\ \emph {et~al.}(2017)\citenamefont {Miller},
  \citenamefont {Steinhoff}, \citenamefont {Pano}, \citenamefont {Klein},
  \citenamefont {Jahnke}, \citenamefont {Holleitner},\ and\ \citenamefont
  {Wurstbauer}}]{Miller.2017}%
  \BibitemOpen
  \bibfield  {author} {\bibinfo {author} {\bibfnamefont {B.}~\bibnamefont
  {Miller}}, \bibinfo {author} {\bibfnamefont {A.}~\bibnamefont {Steinhoff}},
  \bibinfo {author} {\bibfnamefont {B.}~\bibnamefont {Pano}}, \bibinfo {author}
  {\bibfnamefont {J.}~\bibnamefont {Klein}}, \bibinfo {author} {\bibfnamefont
  {F.}~\bibnamefont {Jahnke}}, \bibinfo {author} {\bibfnamefont
  {A.}~\bibnamefont {Holleitner}}, \ and\ \bibinfo {author} {\bibfnamefont
  {U.}~\bibnamefont {Wurstbauer}},\ }\href {\doibase
  10.1021/acs.nanolett.7b01304} {\bibfield  {journal} {\bibinfo  {journal}
  {Nano letters}\ }\textbf {\bibinfo {volume} {17}},\ \bibinfo {pages} {5229}
  (\bibinfo {year} {2017})}\BibitemShut {NoStop}%
\bibitem [{\citenamefont {Robert}\ \emph {et~al.}(2016)\citenamefont {Robert},
  \citenamefont {Lagarde}, \citenamefont {Cadiz}, \citenamefont {Wang},
  \citenamefont {Lassagne}, \citenamefont {Amand}, \citenamefont {Balocchi},
  \citenamefont {Renucci}, \citenamefont {Tongay}, \citenamefont {Urbaszek},\
  and\ \citenamefont {Marie}}]{Robert.2016}%
  \BibitemOpen
  \bibfield  {author} {\bibinfo {author} {\bibfnamefont {C.}~\bibnamefont
  {Robert}}, \bibinfo {author} {\bibfnamefont {D.}~\bibnamefont {Lagarde}},
  \bibinfo {author} {\bibfnamefont {F.}~\bibnamefont {Cadiz}}, \bibinfo
  {author} {\bibfnamefont {G.}~\bibnamefont {Wang}}, \bibinfo {author}
  {\bibfnamefont {B.}~\bibnamefont {Lassagne}}, \bibinfo {author}
  {\bibfnamefont {T.}~\bibnamefont {Amand}}, \bibinfo {author} {\bibfnamefont
  {A.}~\bibnamefont {Balocchi}}, \bibinfo {author} {\bibfnamefont
  {P.}~\bibnamefont {Renucci}}, \bibinfo {author} {\bibfnamefont
  {S.}~\bibnamefont {Tongay}}, \bibinfo {author} {\bibfnamefont
  {B.}~\bibnamefont {Urbaszek}}, \ and\ \bibinfo {author} {\bibfnamefont
  {X.}~\bibnamefont {Marie}},\ }\href@noop {} {\bibfield  {journal} {\bibinfo
  {journal} {Physical Review B}\ }\textbf {\bibinfo {volume} {93}},\ \bibinfo
  {pages} {205423} (\bibinfo {year} {2016})}\BibitemShut {NoStop}%
\bibitem [{\citenamefont {Palummo}\ \emph {et~al.}(2015)\citenamefont
  {Palummo}, \citenamefont {Bernardi},\ and\ \citenamefont
  {Grossman}}]{Palummo.2015}%
  \BibitemOpen
  \bibfield  {author} {\bibinfo {author} {\bibfnamefont {M.}~\bibnamefont
  {Palummo}}, \bibinfo {author} {\bibfnamefont {M.}~\bibnamefont {Bernardi}}, \
  and\ \bibinfo {author} {\bibfnamefont {J.~C.}\ \bibnamefont {Grossman}},\
  }\href {\doibase 10.1021/nl503799t} {\bibfield  {journal} {\bibinfo
  {journal} {Nano letters}\ }\textbf {\bibinfo {volume} {15}},\ \bibinfo
  {pages} {2794} (\bibinfo {year} {2015})}\BibitemShut {NoStop}%
\bibitem [{\citenamefont {Nagler}\ \emph {et~al.}(2017)\citenamefont {Nagler},
  \citenamefont {Plechinger}, \citenamefont {Ballottin}, \citenamefont
  {Mitioglu}, \citenamefont {Meier}, \citenamefont {Paradiso}, \citenamefont
  {Strunk}, \citenamefont {Chernikov}, \citenamefont {Christianen},
  \citenamefont {Sch{\"u}ller},\ and\ \citenamefont {Korn}}]{Nagler.2017}%
  \BibitemOpen
  \bibfield  {author} {\bibinfo {author} {\bibfnamefont {P.}~\bibnamefont
  {Nagler}}, \bibinfo {author} {\bibfnamefont {G.}~\bibnamefont {Plechinger}},
  \bibinfo {author} {\bibfnamefont {M.~V.}\ \bibnamefont {Ballottin}}, \bibinfo
  {author} {\bibfnamefont {A.}~\bibnamefont {Mitioglu}}, \bibinfo {author}
  {\bibfnamefont {S.}~\bibnamefont {Meier}}, \bibinfo {author} {\bibfnamefont
  {N.}~\bibnamefont {Paradiso}}, \bibinfo {author} {\bibfnamefont
  {C.}~\bibnamefont {Strunk}}, \bibinfo {author} {\bibfnamefont
  {A.}~\bibnamefont {Chernikov}}, \bibinfo {author} {\bibfnamefont {P.~C.~M.}\
  \bibnamefont {Christianen}}, \bibinfo {author} {\bibfnamefont
  {C.}~\bibnamefont {Sch{\"u}ller}}, \ and\ \bibinfo {author} {\bibfnamefont
  {T.}~\bibnamefont {Korn}},\ }\href {\doibase 10.1088/2053-1583/aa7352}
  {\bibfield  {journal} {\bibinfo  {journal} {2D Materials}\ }\textbf {\bibinfo
  {volume} {4}},\ \bibinfo {pages} {025112} (\bibinfo {year}
  {2017})}\BibitemShut {NoStop}%
\bibitem [{\citenamefont {Unuchek}\ \emph {et~al.}(2018)\citenamefont
  {Unuchek}, \citenamefont {Ciarrocchi}, \citenamefont {Avsar}, \citenamefont
  {Watanabe}, \citenamefont {Taniguchi},\ and\ \citenamefont
  {Kis}}]{Unuchek.2018}%
  \BibitemOpen
  \bibfield  {author} {\bibinfo {author} {\bibfnamefont {D.}~\bibnamefont
  {Unuchek}}, \bibinfo {author} {\bibfnamefont {A.}~\bibnamefont {Ciarrocchi}},
  \bibinfo {author} {\bibfnamefont {A.}~\bibnamefont {Avsar}}, \bibinfo
  {author} {\bibfnamefont {K.}~\bibnamefont {Watanabe}}, \bibinfo {author}
  {\bibfnamefont {T.}~\bibnamefont {Taniguchi}}, \ and\ \bibinfo {author}
  {\bibfnamefont {A.}~\bibnamefont {Kis}},\ }\href {\doibase
  10.1038/s41586-018-0357-y} {\bibfield  {journal} {\bibinfo  {journal}
  {Nature}\ }\textbf {\bibinfo {volume} {560}},\ \bibinfo {pages} {340}
  (\bibinfo {year} {2018})}\BibitemShut {NoStop}%
\bibitem [{\citenamefont {Kulig}\ \emph {et~al.}(2018)\citenamefont {Kulig},
  \citenamefont {Zipfel}, \citenamefont {Nagler}, \citenamefont {Blanter},
  \citenamefont {Sch{\"u}ller}, \citenamefont {Korn}, \citenamefont {Paradiso},
  \citenamefont {Glazov},\ and\ \citenamefont {Chernikov}}]{Kulig.2018}%
  \BibitemOpen
  \bibfield  {author} {\bibinfo {author} {\bibfnamefont {M.}~\bibnamefont
  {Kulig}}, \bibinfo {author} {\bibfnamefont {J.}~\bibnamefont {Zipfel}},
  \bibinfo {author} {\bibfnamefont {P.}~\bibnamefont {Nagler}}, \bibinfo
  {author} {\bibfnamefont {S.}~\bibnamefont {Blanter}}, \bibinfo {author}
  {\bibfnamefont {C.}~\bibnamefont {Sch{\"u}ller}}, \bibinfo {author}
  {\bibfnamefont {T.}~\bibnamefont {Korn}}, \bibinfo {author} {\bibfnamefont
  {N.}~\bibnamefont {Paradiso}}, \bibinfo {author} {\bibfnamefont {M.~M.}\
  \bibnamefont {Glazov}}, \ and\ \bibinfo {author} {\bibfnamefont
  {A.}~\bibnamefont {Chernikov}},\ }\href {\doibase
  10.1103/PhysRevLett.120.207401} {\bibfield  {journal} {\bibinfo  {journal}
  {Physical review letters}\ }\textbf {\bibinfo {volume} {120}},\ \bibinfo
  {pages} {207401} (\bibinfo {year} {2018})}\BibitemShut {NoStop}%
\bibitem [{\citenamefont {Jauregui}\ \emph {et~al.}(2018)\citenamefont
  {Jauregui}, \citenamefont {Joe}, \citenamefont {Pistunova}, \citenamefont
  {Wild}, \citenamefont {High}, \citenamefont {Zhou}, \citenamefont {Scuri},
  \citenamefont {de~Greve}, \citenamefont {Sushko}, \citenamefont {Yu},
  \citenamefont {Taniguchi}, \citenamefont {Watanabe}, \citenamefont
  {Needleman}, \citenamefont {Lukin}, \citenamefont {Park},\ and\ \citenamefont
  {Kim}}]{Jauregui.2018}%
  \BibitemOpen
  \bibfield  {author} {\bibinfo {author} {\bibfnamefont {L.~A.}\ \bibnamefont
  {Jauregui}}, \bibinfo {author} {\bibfnamefont {A.~Y.}\ \bibnamefont {Joe}},
  \bibinfo {author} {\bibfnamefont {K.}~\bibnamefont {Pistunova}}, \bibinfo
  {author} {\bibfnamefont {D.~S.}\ \bibnamefont {Wild}}, \bibinfo {author}
  {\bibfnamefont {A.~A.}\ \bibnamefont {High}}, \bibinfo {author}
  {\bibfnamefont {Y.}~\bibnamefont {Zhou}}, \bibinfo {author} {\bibfnamefont
  {G.}~\bibnamefont {Scuri}}, \bibinfo {author} {\bibfnamefont
  {K.}~\bibnamefont {de~Greve}}, \bibinfo {author} {\bibfnamefont
  {A.}~\bibnamefont {Sushko}}, \bibinfo {author} {\bibfnamefont {C.-H.}\
  \bibnamefont {Yu}}, \bibinfo {author} {\bibfnamefont {T.}~\bibnamefont
  {Taniguchi}}, \bibinfo {author} {\bibfnamefont {K.}~\bibnamefont {Watanabe}},
  \bibinfo {author} {\bibfnamefont {D.~J.}\ \bibnamefont {Needleman}}, \bibinfo
  {author} {\bibfnamefont {M.~D.}\ \bibnamefont {Lukin}}, \bibinfo {author}
  {\bibfnamefont {H.}~\bibnamefont {Park}}, \ and\ \bibinfo {author}
  {\bibfnamefont {P.}~\bibnamefont {Kim}},\ }\href
  {https://arxiv.org/abs/1812.08691} {\bibfield  {journal} {\bibinfo  {journal}
  {arXiv}\ } (\bibinfo {year} {2018})}\BibitemShut {NoStop}%
\bibitem [{\citenamefont {Gao}\ \emph {et~al.}(2017)\citenamefont {Gao},
  \citenamefont {Yang},\ and\ \citenamefont {Spataru}}]{Gao.2017}%
  \BibitemOpen
  \bibfield  {author} {\bibinfo {author} {\bibfnamefont {S.}~\bibnamefont
  {Gao}}, \bibinfo {author} {\bibfnamefont {L.}~\bibnamefont {Yang}}, \ and\
  \bibinfo {author} {\bibfnamefont {C.~D.}\ \bibnamefont {Spataru}},\ }\href
  {\doibase 10.1021/acs.nanolett.7b04021} {\bibfield  {journal} {\bibinfo
  {journal} {Nano letters}\ }\textbf {\bibinfo {volume} {17}},\ \bibinfo
  {pages} {7809} (\bibinfo {year} {2017})}\BibitemShut {NoStop}%
\bibitem [{\citenamefont {Deilmann}\ and\ \citenamefont
  {Thygesen}(2018)}]{Deilmann.2018}%
  \BibitemOpen
  \bibfield  {author} {\bibinfo {author} {\bibfnamefont {T.}~\bibnamefont
  {Deilmann}}\ and\ \bibinfo {author} {\bibfnamefont {K.~S.}\ \bibnamefont
  {Thygesen}},\ }\href {\doibase 10.1021/acs.nanolett.7b05224} {\bibfield
  {journal} {\bibinfo  {journal} {Nano letters}\ }\textbf {\bibinfo {volume}
  {18}},\ \bibinfo {pages} {1460} (\bibinfo {year} {2018})}\BibitemShut
  {NoStop}%
\bibitem [{\citenamefont {Lucatto}\ \emph {et~al.}()\citenamefont {Lucatto},
  \citenamefont {Koda}, \citenamefont {Bechstedt}, \citenamefont {Marques},\
  and\ \citenamefont {Teles}}]{Lucatto.2019}%
  \BibitemOpen
  \bibfield  {author} {\bibinfo {author} {\bibfnamefont {B.}~\bibnamefont
  {Lucatto}}, \bibinfo {author} {\bibfnamefont {D.~S.}\ \bibnamefont {Koda}},
  \bibinfo {author} {\bibfnamefont {F.}~\bibnamefont {Bechstedt}}, \bibinfo
  {author} {\bibfnamefont {M.}~\bibnamefont {Marques}}, \ and\ \bibinfo
  {author} {\bibfnamefont {L.~K.}\ \bibnamefont {Teles}},\ }\href
  {http://arxiv.org/pdf/1904.10785v1} {\enquote {\bibinfo {title} {Charge qubit
  in van der waals heterostructures},}\ }\BibitemShut {NoStop}%
\bibitem [{SI()}]{SI}%
  \BibitemOpen
  \href@noop {} {\bibinfo  {journal} {For details see supporting information}\
  }\BibitemShut {NoStop}%
\bibitem [{\citenamefont {Castellanos-Gomez}\ \emph {et~al.}(2014)\citenamefont
  {Castellanos-Gomez}, \citenamefont {Buscema}, \citenamefont {Molenaar},
  \citenamefont {Singh}, \citenamefont {Janssen}, \citenamefont {{van der
  Zant}},\ and\ \citenamefont {Steele}}]{CastellanosGomez.2014}%
  \BibitemOpen
\bibfield  {journal} {  }\bibfield  {author} {\bibinfo {author} {\bibfnamefont
  {A.}~\bibnamefont {Castellanos-Gomez}}, \bibinfo {author} {\bibfnamefont
  {M.}~\bibnamefont {Buscema}}, \bibinfo {author} {\bibfnamefont
  {R.}~\bibnamefont {Molenaar}}, \bibinfo {author} {\bibfnamefont
  {V.}~\bibnamefont {Singh}}, \bibinfo {author} {\bibfnamefont
  {L.}~\bibnamefont {Janssen}}, \bibinfo {author} {\bibfnamefont {H.~S.~J.}\
  \bibnamefont {{van der Zant}}}, \ and\ \bibinfo {author} {\bibfnamefont
  {G.~A.}\ \bibnamefont {Steele}},\ }\href {\doibase
  10.1088/2053-1583/1/1/011002} {\bibfield  {journal} {\bibinfo  {journal} {2D
  Materials}\ }\textbf {\bibinfo {volume} {1}},\ \bibinfo {pages} {011002}
  (\bibinfo {year} {2014})}\BibitemShut {NoStop}%
\bibitem [{\citenamefont {Okada}\ \emph {et~al.}(2018)\citenamefont {Okada},
  \citenamefont {Kutana}, \citenamefont {Kureishi}, \citenamefont {Kobayashi},
  \citenamefont {Saito}, \citenamefont {Saito}, \citenamefont {Watanabe},
  \citenamefont {Taniguchi}, \citenamefont {Gupta}, \citenamefont {Miyata},
  \citenamefont {Yakobson}, \citenamefont {Shinohara},\ and\ \citenamefont
  {Kitaura}}]{Okada.2018}%
  \BibitemOpen
  \bibfield  {author} {\bibinfo {author} {\bibfnamefont {M.}~\bibnamefont
  {Okada}}, \bibinfo {author} {\bibfnamefont {A.}~\bibnamefont {Kutana}},
  \bibinfo {author} {\bibfnamefont {Y.}~\bibnamefont {Kureishi}}, \bibinfo
  {author} {\bibfnamefont {Y.}~\bibnamefont {Kobayashi}}, \bibinfo {author}
  {\bibfnamefont {Y.}~\bibnamefont {Saito}}, \bibinfo {author} {\bibfnamefont
  {T.}~\bibnamefont {Saito}}, \bibinfo {author} {\bibfnamefont
  {K.}~\bibnamefont {Watanabe}}, \bibinfo {author} {\bibfnamefont
  {T.}~\bibnamefont {Taniguchi}}, \bibinfo {author} {\bibfnamefont
  {S.}~\bibnamefont {Gupta}}, \bibinfo {author} {\bibfnamefont
  {Y.}~\bibnamefont {Miyata}}, \bibinfo {author} {\bibfnamefont {B.~I.}\
  \bibnamefont {Yakobson}}, \bibinfo {author} {\bibfnamefont {H.}~\bibnamefont
  {Shinohara}}, \ and\ \bibinfo {author} {\bibfnamefont {R.}~\bibnamefont
  {Kitaura}},\ }\href {\doibase 10.1021/acsnano.7b08253} {\bibfield  {journal}
  {\bibinfo  {journal} {ACS nano}\ }\textbf {\bibinfo {volume} {12}},\ \bibinfo
  {pages} {2498} (\bibinfo {year} {2018})}\BibitemShut {NoStop}%
\bibitem [{\citenamefont {Wierzbowski}\ \emph {et~al.}(2017)\citenamefont
  {Wierzbowski}, \citenamefont {Klein}, \citenamefont {Sigger}, \citenamefont
  {Straubinger}, \citenamefont {Kremser}, \citenamefont {Taniguchi},
  \citenamefont {Watanabe}, \citenamefont {Wurstbauer}, \citenamefont
  {Holleitner}, \citenamefont {Kaniber}, \citenamefont {M{\"u}ller},\ and\
  \citenamefont {Finley}}]{Wierzbowski.2017}%
  \BibitemOpen
  \bibfield  {author} {\bibinfo {author} {\bibfnamefont {J.}~\bibnamefont
  {Wierzbowski}}, \bibinfo {author} {\bibfnamefont {J.}~\bibnamefont {Klein}},
  \bibinfo {author} {\bibfnamefont {F.}~\bibnamefont {Sigger}}, \bibinfo
  {author} {\bibfnamefont {C.}~\bibnamefont {Straubinger}}, \bibinfo {author}
  {\bibfnamefont {M.}~\bibnamefont {Kremser}}, \bibinfo {author} {\bibfnamefont
  {T.}~\bibnamefont {Taniguchi}}, \bibinfo {author} {\bibfnamefont
  {K.}~\bibnamefont {Watanabe}}, \bibinfo {author} {\bibfnamefont
  {U.}~\bibnamefont {Wurstbauer}}, \bibinfo {author} {\bibfnamefont {A.~W.}\
  \bibnamefont {Holleitner}}, \bibinfo {author} {\bibfnamefont
  {M.}~\bibnamefont {Kaniber}}, \bibinfo {author} {\bibfnamefont
  {K.}~\bibnamefont {M{\"u}ller}}, \ and\ \bibinfo {author} {\bibfnamefont
  {J.~J.}\ \bibnamefont {Finley}},\ }\href {\doibase
  10.1038/s41598-017-09739-4} {\bibfield  {journal} {\bibinfo  {journal}
  {Scientific reports}\ }\textbf {\bibinfo {volume} {7}},\ \bibinfo {pages}
  {12383} (\bibinfo {year} {2017})}\BibitemShut {NoStop}%
\bibitem [{\citenamefont {Cadiz}\ \emph {et~al.}(2017)\citenamefont {Cadiz},
  \citenamefont {Courtade}, \citenamefont {Robert}, \citenamefont {Wang},
  \citenamefont {Shen}, \citenamefont {Cai}, \citenamefont {Taniguchi},
  \citenamefont {Watanabe}, \citenamefont {Carrere}, \citenamefont {Lagarde},
  \citenamefont {Manca}, \citenamefont {Amand}, \citenamefont {Renucci},
  \citenamefont {Tongay}, \citenamefont {Marie},\ and\ \citenamefont
  {Urbaszek}}]{Cadiz.2017}%
  \BibitemOpen
  \bibfield  {author} {\bibinfo {author} {\bibfnamefont {F.}~\bibnamefont
  {Cadiz}}, \bibinfo {author} {\bibfnamefont {E.}~\bibnamefont {Courtade}},
  \bibinfo {author} {\bibfnamefont {C.}~\bibnamefont {Robert}}, \bibinfo
  {author} {\bibfnamefont {G.}~\bibnamefont {Wang}}, \bibinfo {author}
  {\bibfnamefont {Y.}~\bibnamefont {Shen}}, \bibinfo {author} {\bibfnamefont
  {H.}~\bibnamefont {Cai}}, \bibinfo {author} {\bibfnamefont {T.}~\bibnamefont
  {Taniguchi}}, \bibinfo {author} {\bibfnamefont {K.}~\bibnamefont {Watanabe}},
  \bibinfo {author} {\bibfnamefont {H.}~\bibnamefont {Carrere}}, \bibinfo
  {author} {\bibfnamefont {D.}~\bibnamefont {Lagarde}}, \bibinfo {author}
  {\bibfnamefont {M.}~\bibnamefont {Manca}}, \bibinfo {author} {\bibfnamefont
  {T.}~\bibnamefont {Amand}}, \bibinfo {author} {\bibfnamefont
  {P.}~\bibnamefont {Renucci}}, \bibinfo {author} {\bibfnamefont
  {S.}~\bibnamefont {Tongay}}, \bibinfo {author} {\bibfnamefont
  {X.}~\bibnamefont {Marie}}, \ and\ \bibinfo {author} {\bibfnamefont
  {B.}~\bibnamefont {Urbaszek}},\ }\href@noop {} {\bibfield  {journal}
  {\bibinfo  {journal} {Physical Review X}\ }\textbf {\bibinfo {volume} {7}},\
  \bibinfo {pages} {021026} (\bibinfo {year} {2017})}\BibitemShut {NoStop}%
\bibitem [{\citenamefont {Nayak}\ \emph {et~al.}(2017)\citenamefont {Nayak},
  \citenamefont {Horbatenko}, \citenamefont {Ahn}, \citenamefont {Kim},
  \citenamefont {Lee}, \citenamefont {Ma}, \citenamefont {Jang}, \citenamefont
  {Lim}, \citenamefont {Kim}, \citenamefont {Ryu}, \citenamefont {Cheong},
  \citenamefont {Park},\ and\ \citenamefont {Shin}}]{Nayak.2017}%
  \BibitemOpen
  \bibfield  {author} {\bibinfo {author} {\bibfnamefont {P.~K.}\ \bibnamefont
  {Nayak}}, \bibinfo {author} {\bibfnamefont {Y.}~\bibnamefont {Horbatenko}},
  \bibinfo {author} {\bibfnamefont {S.}~\bibnamefont {Ahn}}, \bibinfo {author}
  {\bibfnamefont {G.}~\bibnamefont {Kim}}, \bibinfo {author} {\bibfnamefont
  {J.-U.}\ \bibnamefont {Lee}}, \bibinfo {author} {\bibfnamefont {K.~Y.}\
  \bibnamefont {Ma}}, \bibinfo {author} {\bibfnamefont {A.-R.}\ \bibnamefont
  {Jang}}, \bibinfo {author} {\bibfnamefont {H.}~\bibnamefont {Lim}}, \bibinfo
  {author} {\bibfnamefont {D.}~\bibnamefont {Kim}}, \bibinfo {author}
  {\bibfnamefont {S.}~\bibnamefont {Ryu}}, \bibinfo {author} {\bibfnamefont
  {H.}~\bibnamefont {Cheong}}, \bibinfo {author} {\bibfnamefont
  {N.}~\bibnamefont {Park}}, \ and\ \bibinfo {author} {\bibfnamefont {H.~S.}\
  \bibnamefont {Shin}},\ }\href {\doibase 10.1021/acsnano.7b00640} {\bibfield
  {journal} {\bibinfo  {journal} {ACS nano}\ }\textbf {\bibinfo {volume}
  {11}},\ \bibinfo {pages} {4041} (\bibinfo {year} {2017})}\BibitemShut
  {NoStop}%
\bibitem [{\citenamefont {Varshni}(1967)}]{Varshni.1967}%
  \BibitemOpen
  \bibfield  {author} {\bibinfo {author} {\bibfnamefont {Y.~P.}\ \bibnamefont
  {Varshni}},\ }\href {\doibase 10.1016/0031-8914(67)90062-6} {\bibfield
  {journal} {\bibinfo  {journal} {Physica}\ }\textbf {\bibinfo {volume} {34}},\
  \bibinfo {pages} {149} (\bibinfo {year} {1967})}\BibitemShut {NoStop}%
\bibitem [{\citenamefont {Yu}\ \emph {et~al.}(2015)\citenamefont {Yu},
  \citenamefont {Wang}, \citenamefont {Tong}, \citenamefont {Xu},\ and\
  \citenamefont {Yao}}]{Yu.2015}%
  \BibitemOpen
  \bibfield  {author} {\bibinfo {author} {\bibfnamefont {H.}~\bibnamefont
  {Yu}}, \bibinfo {author} {\bibfnamefont {Y.}~\bibnamefont {Wang}}, \bibinfo
  {author} {\bibfnamefont {Q.}~\bibnamefont {Tong}}, \bibinfo {author}
  {\bibfnamefont {X.}~\bibnamefont {Xu}}, \ and\ \bibinfo {author}
  {\bibfnamefont {W.}~\bibnamefont {Yao}},\ }\href {\doibase
  10.1103/PhysRevLett.115.187002} {\bibfield  {journal} {\bibinfo  {journal}
  {Physical review letters}\ }\textbf {\bibinfo {volume} {115}},\ \bibinfo
  {pages} {187002} (\bibinfo {year} {2015})}\BibitemShut {NoStop}%
\bibitem [{\citenamefont {Ge}\ and\ \citenamefont {Liu}(2013)}]{Ge.2013}%
  \BibitemOpen
  \bibfield  {author} {\bibinfo {author} {\bibfnamefont {Y.}~\bibnamefont
  {Ge}}\ and\ \bibinfo {author} {\bibfnamefont {A.~Y.}\ \bibnamefont {Liu}},\
  }\href {\doibase 10.1103/PhysRevB.87.241408} {\bibfield  {journal} {\bibinfo
  {journal} {Physical Review B}\ }\textbf {\bibinfo {volume} {87}},\ \bibinfo
  {pages} {241408(R)} (\bibinfo {year} {2013})}\BibitemShut {NoStop}%
\end{thebibliography}%

\end{document}